\documentclass[traditabstract]{aa} 
\usepackage{graphicx}
\usepackage{txfonts}
\usepackage{natbib}

\def\champp {CHAMP$^+$}
\def\micron {$\mu$m}

\def\Lsol {$L_\odot$}
\def\Msol {$M_\odot$}

\def\HH {H$_{2}$}
\def\Water {H$_{2}$O}

\def\TWCO {$^{12}$CO}

\def\CXVIIIO {C$^{18}$O}

\def\kms {km~s$^{-1}$}
\def\vlsr {V$_{\rm LSR}$}

\def\aap{A\&A}                
\def\aapr{A\&A~Rev.}          
\def\aj{AJ}                   
\def\apj{ApJ}                 
\def\apjl{ApJL}                
\def\apjs{ApJS}               
\def\mnras{MNRAS}             
\def\nat{Nature}              
\begin{document}
\title{Mid-$J$ CO emission from the Orion BN/KL explosive outflow}

   \author{T.-C. Peng\inst{1,2,3}
          \and L. A. Zapata\inst{4}
          \and F. Wyrowski\inst{1}
          \and R. G{\"u}sten\inst{1}
          \and K. M. Menten\inst{1}
          }

   \institute{Max-Planck-Institut f\"ur Radioastronomie (MPIfR),
              Auf dem H\"ugel 69, 53121 Bonn, Germany 
              \and Univ. Bordeaux, LAB, UMR 5804, F-33270, Floirac, France 
              \and CNRS, LAB, UMR 5804, F-33270, Floirac, France \\
              \email{Tzu-Cheng.Peng@obs.u-bordeaux1.fr}
              \and Centro de Radioastronom\'\i a y Astrof\'\i sica,
Universidad Nacional Aut\'onoma de M\'exico, Morelia 58090, M\'exico
             }

\date{Received; accepted}
\titlerunning{APEX-\champp\ observations of the Orion BN/KL explosive outflow}
\authorrunning{Peng et al.}

 
  
  \abstract{High spatial resolution low-$J$ \TWCO\ observations have shown that the wide-angle outflow seen in the Orion BN/KL region correlates with the famous \HH\ fingers. Recently, high-resolution large-scale mappings of mid- and higher-$J$ CO emissions have been reported toward the Orion molecular cloud 1 core region using the APEX telescope. Therefore, it is of interest to investigate this outflow in the higher-$J$ \TWCO\ emission, which is likely excited by shocks. The observations were carried out using the dual-color heterodyne array \champp\ on the APEX telescope. The images of the Orion BN/KL region were obtained in the \TWCO\ $J$=6--5 and $J$=7--6 transitions with angular resolutions of $8\farcs6$ and $7\farcs4$, respectively. The results show a good agreement between our higher-$J$ \TWCO\ emission and SMA low-$J$ \TWCO\ data, which indicates that this wide-angle outflow in Orion BN/KL is likely the result of an explosive event that is related to the runaway objects from a dynamically decayed multiple system. From our observations, we estimate that the kinetic energy of this explosive outflow is about $1-2\times10^{47}$ erg. In addition, a scenario has been proposed where part of the outflow is decelerated and absorbed in the cloud to explain the lack of CO bullets in the southern part of BN/KL, which in turn induces the methanol masers seen in this region.}


   \maketitle
%

\section{Introduction}

Orion BN/KL \citep{Becklin1967,Kleinmann1967}, the most luminous source within the Orion molecular cloud 1 (OMC-1), has an infrared (IR) luminosity of $\sim 10^{5}$ \Lsol\ and has been intensively observed in numerous molecular lines and also at various wavelengths \citep[][]{O'Dell2008}. Orion BN/KL harbors the eponymous molecular ``hot core'' (HC) with a rich and complex chemistry \citep[e.g.,][]{Blake1987}. In addition, shock and outflow phenomena are also widely present in the Orion BN/KL region. Near-infrared (NIR) observations showed \HH\ outflows in the form of bullets or fingers \citep{Taylor1984,Allen1993} at a wide angle along a northwest-southeast axis. In radio observations, the powerful outflow was also seen in the high-velocity \TWCO\ line wings \citep{Kwan1976} together with some hydroxyl (OH) masers that trace the shock fronts \citep{Norris1984,Cohen2006}. Another outflow with lower velocities was detected through proper motions of \Water\ masers \citep{Genzel1981,Gaume1998} accompanied with SiO masers near the IR source IRc2 \citep{Greenhill1998,Doeleman1999}, and is closely associated with source $I$ \citep[][]{Menten1995,Reid2007}.

The origin of the wide-angle high-velocity outflow and northeast-southwest low-velocity outflow has been frequently discussed, and new debates have begun with updated observations. \citet{Zapata2009} reported explosive \TWCO\ bullets in high spatial resolution associated with the \HH\ fingers, and indicated that this outflow is caused by disintegration of a massive stellar system that happened about 500 years ago. Evidence of a close dynamical interaction of a young stellar system is also supported by the proper motions of the Orion BN source, radio source $I$, and/or source $n$, which may be runaway objects from a dynamically decayed multiple system \citep{Nissen2012,Goddi2011,Bally2011,Rodriguez2005,Gomez2005}. Such an event could also produce the energy needed for driving the large-scale outflows in the Orion BN/KL region \citep{Bally2005} and may indeed be responsible for most of the luminosity in the region because unlike other hot molecular cores, the Orion HC (mm/submm continuum emission peak) does not appear to contain an embedded massive protostellar object capable of producing the necessary energy \citep{Zapata2011}. In this paper, we present the results of the new \TWCO\ $J$=6--5 and $J$=7--6 observations toward the Orion BN/KL explosive outflow. 





\section{\label{obs}Observations}

The observations were carried out in 2007 November and 2008 July with the 12 meter APEX telescope on Llano de Chajnantor in Chile \citep{Gusten2006}\footnote{This publication is based on data acquired with the Atacama Pathfinder Experiment (APEX). APEX is a collaboration between the Max-Planck-Institut f\"ur Radioastronomie, the European Southern Observatory, and the Onsala Space Observatory.}, and were part of the large-scale on-the-fly (OTF) mapping in the OMC-1 core region \citep[for more observational details, see][]{Peng2012}. The observations of the \TWCO\ $J$=6--5 (691.5 GHz) and $J$=7--6 (806.7 GHz) lines were performed with the dual-color heterodyne array \champp\ \citep{Kasemann2006,Gusten2008}, operating $2\times7$ elements simultaneously in the 450 and 350 \micron\ atmospheric windows. The MPIfR Array Correlator Spectrometer (MACS) and the fast Fourier transform spectrometers \citep[FFTS,][]{Klein2006} were used as backends. Spectral resolutions were $\sim$ 0.42 km s$^{-1}$ for the MACS and $\sim$ 0.05 km s$^{-1}$ for the FFTS. The angular resolutions of \champp\ are $7\farcs4$ at 806.7 GHz and $8\farcs6$ at 691.5 GHz. The maps were centered on the Orion BN source ($\alpha,\delta_{J2000}= 05^{\rm{h}}35^{\rm{m}}14\fs16$, $-05\degr22\arcmin21\farcs5$). During the observations, position switching to an emission-free reference position at offsets ($-500$\arcsec, 0\arcsec) was used. Mira ($o$ Ceti) was used as a line-pointing source in the \TWCO\ $J$=6--5 line, and the overall pointing accuracy is estimated to be $\leq4\arcsec$. All spectra were converted to the main beam brightness temperature unit, $T_{\rm MB}=T_{\rm A}^{*}/\eta_{\rm MB}$ ($\eta_{\rm MB}$ is 0.47 and 0.45 for \TWCO\ $J$=6--5 and $J$=7--6, respectively), using a forward efficiency of 0.95 and beam coupling efficiencies as determined toward Jupiter. All data were reduced using the standard procedures in the GILDAS package\footnote{http://www.iram.fr/IRAMFR/GILDAS/}.
\begin{figure*}
 \centering
 \includegraphics[angle=270,width=0.88\textwidth]{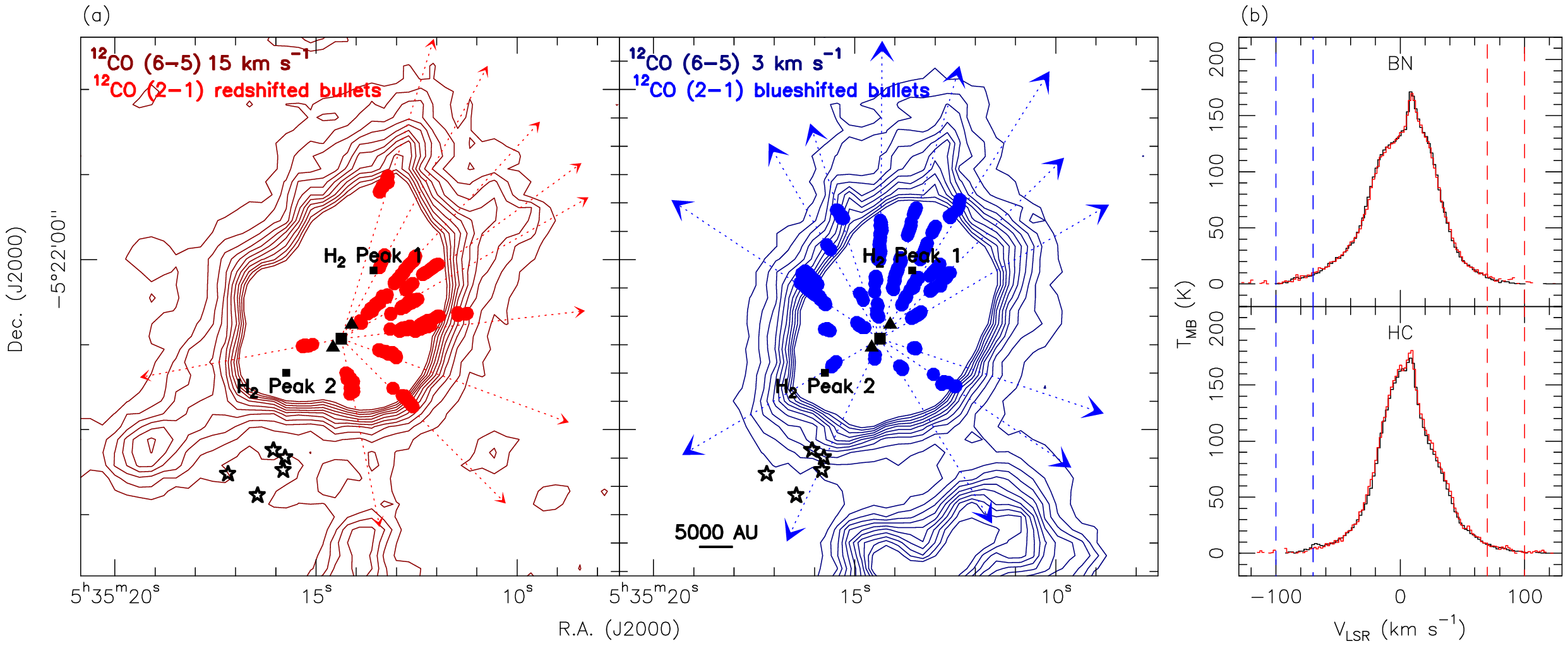}
 \caption{a) Redshifted \TWCO\ $J$=6--5 emission (left panel) at the 15 \kms\ channel ($\Delta V=1$ \kms) and the blueshifted \TWCO\ emission (right panel) at the 3 \kms\ channel ($\Delta V=1$ \kms) overlaid with the SMA \TWCO\ $J$=2--1 bullets taken from \citet{Zapata2009}. The \TWCO\ $J$=6--5 emission is plotted in dark red and dark blue contours running from 8 to 48 K \kms\ in steps of 4 k \kms (1 $\sigma$), and the contours with intensities higher than 48 K \kms\ were skipped for clarity. The redshifted (from $+$40 to $+$120 \kms) and blueshifted (from --30 to --110 \kms) \TWCO\ $J$=2--1 emission peaks at different velocities are plotted as red and blue dots. The squares represent the positions of the outflow explosion center and the \HH\ emission peak 1 and 2. The triangles represent the positions of the Orion HC (south) and BN (north). The stars mark the positions of the five Trapezium stars ($\theta^1$ Ori A, B, C, D, and E). The positions of the \HH\ emission peak 1 and 2 are marked as squares. Note that the emission in the southwest of the Trapezium stars comes from the OMC-1S region. b) \TWCO\ $J$=6--5 (black) and $J$=7--6 (red) spectra toward Orion BN (upper panel) and the HC (lower panel). Dashed lines indicate the integrated velocity ranges shown in Figs. \ref{kl-outflows2} and \ref{orion-kl-outflow-sma-2}. \label{orion-kl-outflow-sma-1}}
 \end{figure*}

\section{Results and discussion}

 \begin{figure}
 \centering
 \includegraphics[angle=270,width=0.30\textwidth]{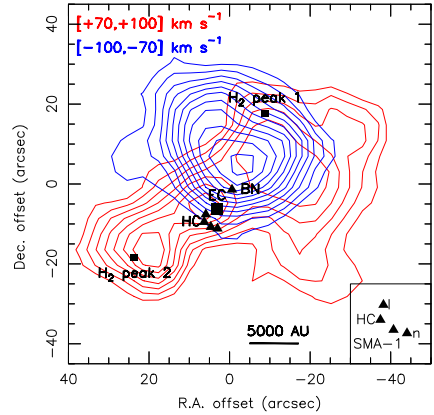}
 \caption{Image of the very high velocity emission in the \TWCO\ $J$=7--6 transition (smoothed by a Gaussian profile with a width of $16\farcs4$). Red contours are plotted from 30 to 90 K \kms\ in steps of 10 K \kms\ (2 $\sigma$ level), and blue contours are plotted from 30 to 130 K \kms\ in steps of 10 K \kms\ (2 $\sigma$ level). The positions of source $I$, the HC, SMA-1, and source $n$ are represented as triangles, and their relative positions are indicated in the bottom-right panel. The squares represent the positions of the outflow explosion center (EC) and the \HH\ emission peak 1 and 2. \label{kl-outflows2}}
 \end{figure}

 \begin{figure*}
 \centering
 \includegraphics[angle=270,width=0.9\textwidth]{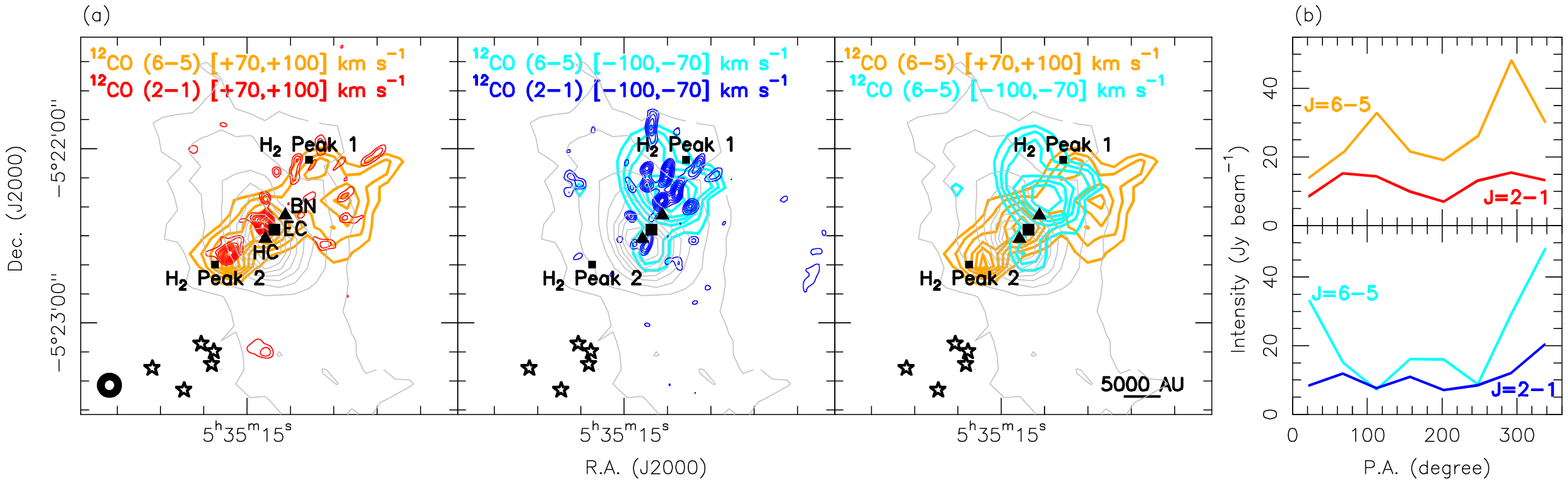}
 \caption{a) Very high velocity outflow emission in the \TWCO\ $J$=6--5 transition overlaid with the SMA \TWCO\ $J$=2--1 emission \citep[$\theta_{\rm syn}=3\farcs3\times3\farcs1$, P.A.=--14\degr,][]{Zapata2009} and the \HH\ column density \citep{Peng2012}. The left panel shows the redshifted \TWCO\ emission integrated from $+$70 to $+$100 \kms\ (\vlsr), and the middle panel shows the blueshifted \TWCO\ emission integrated from $-100$ to $-70$ \kms. The \TWCO\ $J$=6--5 emission is plotted in orange and light blue contours running from $10\%$ to $90\%$ in steps of $10\%$ of the peak intensity. The \TWCO\ $J$=2--1 emission is plotted in thick red and blue contours from $30\%$ to $90\%$ in steps of $20\%$ of the peak intensity. The \HH\ column density \citep[derived from CO isotopologs,][]{Peng2012} is plotted in gray contours running from $10\%$ to $90\%$ in steps of $10\%$ of the peak density ($5.8\times10^{23}$ cm$^{-2}$). The right panel shows the superposition of the high-velocity \TWCO\ emission from the left and middle panels, and the beam sizes of APEX and SMA are indicated in the first panel. The squares represent the positions of the outflow explosion center (EC) and the \HH\ emission peak 1 and 2. The triangles represent the positions of the Orion HC (south) and BN (north). The stars mark the positions of the five Trapezium stars ($\theta^1$ Ori A, B, C, D, and E). b) The corresponding mean intensities (redshifted and blueshifted) are measured in different position angles (in steps of 45\degr). Note that the SMA \TWCO\ $J$=2--1 intensities were multiplied by 3. \label{orion-kl-outflow-sma-2}}
 \end{figure*}

In Figure \ref{orion-kl-outflow-sma-1} a, the faint high-velocity \TWCO\ $J$=6--5 emission around the BN/KL core region shows a clear footprint from the finger-like structure revealed by the SMA \TWCO\ $J$=2--1 bullets \citep[][]{Zapata2009}, especially in the northwest (redshifted components) and north (blueshifted components). In addition, Figure \ref{kl-outflows2} shows the very high velocity (up to $\pm100$ \kms) components in the \TWCO\ $J$=7--6 emission, and the \TWCO\ outflows are clearly associated with the \HH\ emission peak 1 and 2, the two strongest peaks of vibrationally excited \HH\ emission \citep{Sugai1994,Beckwith1978}. The \TWCO\ $J$=7--6 and $J$=6--5 spectra toward the Orion HC and BN are shown in Figure \ref{orion-kl-outflow-sma-1} b, where the extended line wings up to $\pm100$ \kms\ are clearly visible. This very high velocity outflow has been observed in different \TWCO\ transitions before \citep[see, e.g.,][]{Wilson2001,Rodriguez-Franco1999,Schulz1995,Schmid-Burgk1989}, and the morphology of the high-velocity \TWCO\ $J$=7--6 emission is similar to the result obtained by \citet{Rodriguez-Franco1999}, where the high-speed blueshifted and redshifted bullets seem to be associated with and preceded by the \HH\ emission peak 1 and 2, respectively. However, the upper limit of the \HH\ outflow radial velocity is $\sim 34-40$ \kms\ \citep[][]{Nissen2007}, which is only half the velocity of the high-speed \TWCO\ bullets. Another noteworthy similarity between the \HH\ fingers and \TWCO\ outflows is that the wide-angle distribution in the northwestern part is presented both in the \HH\ observations and our data.


\begin{table}
\caption{Orion BN/KL explosive outflow parameters}             
\label{table1}      
\centering                          
\begin{tabular}{lcccccc}        
\hline\hline                 
Component\tablefootmark{a} & $f$\tablefootmark{b}  & $A_{\rm APEX}$\tablefootmark{b}     & $A_{\rm SMA}$\tablefootmark{b}       & $M_{\rm outflow}$   &  Energy\tablefootmark{c}   \\
                           & ($\%$)                & \multicolumn{2}{c}{(10$^{-3}$ $\rm pc^2$)} & (\Msol)             & ($10^{46}$ erg)     \\
\hline                        
$[-100,-70]$               &  $\sim$22             & 6.5 & 1.4  &  0.20--0.24        & 1.8--2.2   \\
$[-70,-30]$                &  $\sim$25             & 7.7 & 1.9  &  1.71--2.09        & 6.1--7.5   \\
$[+30,+70]$                &  $\sim$26             & 8.9 & 2.3  &  2.61--3.18        & 4.1--5.1   \\
$[+70,+100]$               &  $\sim$14             & 8.8 & 1.2  &  0.41--0.50        & 2.3--2.8   \\

\hline                                   
\end{tabular}

\tablefoot{
\tablefoottext{a}{Integrated velocity ranges in \vlsr.}
\tablefoottext{b}{The filling factors were estimated by comparing the \TWCO\ $J$=6--5 emission sizes ($A_{\rm APEX}$) with those of the SMA \TWCO\ $J$=2--1 emission ($A_{\rm SMA}$) at the contours of 20\% of the peak intensities.}
\tablefoottext{c}{The outflow energy ($\frac{1}{2}M_{\rm outflow}V^2$) was calculated by using the median velocities for the redshifted and blueshifted components.}
}

\end{table}

The very high velocity \TWCO\ $J$=2--1 emission (Fig. \ref{orion-kl-outflow-sma-2} a) also shows clear emission counterparts in the APEX \TWCO\ $J$=6--5 emission, where the redshifted emission lies mostly in the NW-SE direction and the blueshifted one lies mostly in the N-S direction. The average intensities over different position angles (Fig. \ref{orion-kl-outflow-sma-2} b) also indicate the similarity between SMA \TWCO\ $J$=2--1 and APEX $J$=6--5 lines. Interestingly, the dense ridge peaking at the Orion HC is located to the east of the explosion center.

The column density of the outflow can be estimated using
\begin{equation}
N({\rm ^{12}CO})=\frac{3kQ_{\rm rot}}{8\pi^{3}\nu S\mu^{2}f}e^{\frac{E_{\rm up}}{kT_{\rm ex}}}\int T_{\rm MB}dV,
\end{equation}
assuming optically thin \TWCO\ $J$=6--5 emission at high velocities. The optical depth of \TWCO\ $J$=6--5 decreases to about 0.1 at 20 \kms, estimated from the \CXVIIIO\ (6-5)/\TWCO\ (6-5) ratios \citep{Peng2012} with a [\TWCO]/[\CXVIIIO] ratio of 490 \citep{Boreiko1996,Wilson1992}. $Q_{\rm rot}$ is the rotational partition function, $\mu$ the dipole moment of \TWCO, $S$ the line strength, and $T_{\rm ex}$ the excitation temperature. The upper level energy $E_{\rm up}/k$ for \TWCO\ $J$=6--5 is 116 K. The beam filling factor $f$ was estimated by comparing the outflow sizes (at the contours with 20\% of peak intensities) of the \TWCO\ $J$=2--1 interferometry data with the APEX \TWCO\ $J$=6--5 emission sizes, given that these bullets were resolved by the SMA beam ($3\farcs28\times3\farcs12$). The excitation temperature $T_{\rm ex}$ of about 60 K was estimated from the ratio between $J$=2--1 and $J$=6--5 in the LTE condition. The outflow mass and kinetic energy then can be calculated by adopting a [\TWCO]/[\HH] abundance ratio of 8 $\times\ 10^{-5}$ in OMC-1 \citep{Wilson1992}. The derived parameters of this outflow are listed in Table \ref{table1}. The total kinetic energy associated with this explosive outflow in Orion BN/KL is about $1-2\times10^{47}$ erg, close to the value derived by \citet{Kwan1976} with a revised distance to the Orion Nebula \citep[414$\pm$7 pc,][]{Menten2007}. The outflow gas with low velocities ($-30<V_{\rm LSR}<30$ \kms) is likely contaminated by ambient gas and was excluded from the calculation (the filling factor is also difficult to estimate due to the filtering effect of the SMA data). It is likely that the low-velocity gas only contributes a small fraction of the total outflow mass and kinetic energy. Additionally, the good agreement between the high-resolution low-$J$ \TWCO\ observations \citep{Zapata2009} and our higher-$J$ \TWCO\ data indicates a close spatial correlation between the IR \HH\ fingers and the molecular outflows, which are likely driven by an explosive event instead of the outflows from a single young stellar object or radio source, i.e., SMA-1 or IRc2. Nevertheless, if this explosive outflow is a superposition of several flows emanating from protostellar objects in BN/KL, they are likely distributed uniformly in position angle and will show little asymmetry in redshifted and blueshifted components, which are ruled out by our results.

\begin{figure}
 \centering
 \includegraphics[angle=270,width=0.42\textwidth]{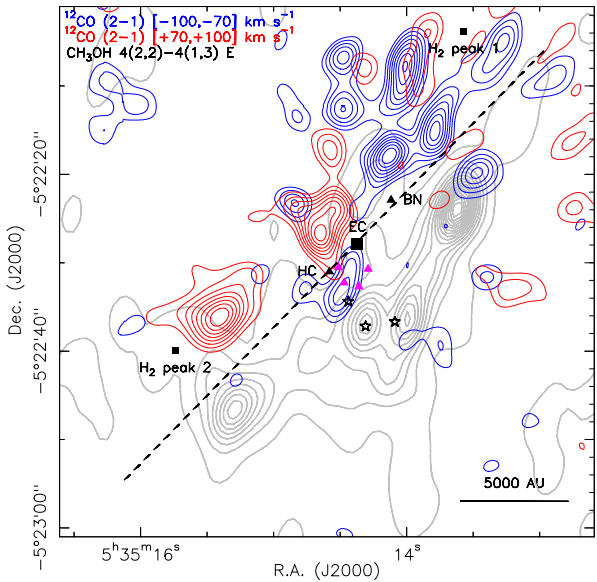}
 \caption{SMA high-velocity \TWCO\ bullets overlaid with the JVLA methanol maser emission (K. M. Menten, priv. comm.). The red and blue \TWCO\ emission contours are the same as shown in Figs. \ref{orion-kl-outflow-sma-1} and \ref{orion-kl-outflow-sma-2}. The CH$_3$OH $4_{2,2}-4_{1,3}$ E maser emission is shown in grey contours plotted 0.05 to 3.05 Jy beam$^{-1}$ in step of 0.2 Jy beam$^{-1}$. The black dashed line indicates the possible symmetric distribution between the \TWCO\ and CH$_3$OH emission, where most of the high-velocity \TWCO\ bullets are likely absorbed below this line. The positions of the explosion center (EC), source BN, HC, and \HH\ peak 1 and 2 are marked. Magenta triangles represent the positions of source $I$, SMA-1, source $n$, and IRc7 (east to west). The three strongest methyl formate emission peaks are marked in black stars \citep{Favre2011}. \label{Fig-large-scale}}
 \end{figure}

\citet{Zapata2009,Zapata2011} showed that the \TWCO\ bullets are less present in the southern and southeastern directions (see also Fig. \ref{orion-kl-outflow-sma-2} b). Because the N-S dense ridge in Orion BN/KL has a velocity gradient where the northern part (with respect to the HC) has a LSR velocity of $\gtrsim9$ \kms, and $\lesssim8$ \kms\ for the denser southern part \citep{Peng2012}, it is likely that most of the bullets are decelerated and eventually absorbed in the southern part of the cloud. This scenario is supported by the detection of methanol masers (e.g., the $4_{2,2}-4_{1,3}$ E line shown in Fig. \ref{Fig-large-scale}), which are located mostly in the southern part of the ridge and appear to have formed behind the shock fronts \citep{Johnston1992}. Therefore, it is possible that the shocks due to this explosive outflow are absorbed in the southern part of Orion BN/KL and induce the methanol masers. There may be temperature differences in the cloud that affect the formation and/or distribution of some molecules. This scenario may help to explain the long-debated differentiation between oxygen- and nitrogen-bearing species \citep[see, e.g.,][]{Blake1987,Rodgers2001,Liu2002,Friedel2008}, namely that its main cause may be related to shocks.

Furthermore, if one-forth of the bullets ($\sim2.5\times10^{46}$ erg) are absorbed by the cloud in 500 years, the resulting luminosity ($L\sim E/t$) is $\gtrsim400$ \Lsol, where the time scale for absorption and re-radiation is shorter than 500 years because the ejecta would take time to reach the outer part of the cloud. The total bolometric luminosity for BN/KL is about $7\times10^4$ \Lsol\ \citep{Gezari1998} assuming a distance of 414 pc, and therefore the explosive outflow may not be energetic enough to illuminate the whole BN/KL region (other stellar objects may also contribute). However, IRc2 with a luminosity of just 1000$\pm$500 \Lsol\ and the other less luminous infrared sources in BN/KL \citep{Gezari1998} may be illuminated by this outflow.

\section{Summary}

We detected very high velocity (up to $\pm$100 \kms) outflows in the \TWCO\ $J$=6--5 and $J$=7--6 transitions in the Orion BN/KL region. The morphology of our \TWCO\ emission agrees well with the high-resolution SMA observations, where the molecular outflow is driven by an explosive event and is related to the IR \HH\ fingers seen on a larger scale in Orion BN/KL. The kinetic energy of this explosive outflow calculated from the high-velocity CO bullets is estimated to be about $1-2\times10^{47}$ erg. It is still not clear what caused this explosion, but it is likely related to the runaway objects from a dynamically decayed multiple system in Orion BN/KL. We also proposed a scenario in which the methanol masers are induced by the same outflow at which most of the shocks are decelerated and absorbed in the southern part of BN/KL. 



\begin{acknowledgements}
We thank the APEX staff for the support during the observations and the anonymous referee for improving the paper. This work was supported by the International Max Planck Research School (IMPRS) for Astronomy and Astrophysics at the Universities of Bonn and Cologne.
\end{acknowledgements}

\end{document}